\documentclass{appolb}

\usepackage{graphicx}
\usepackage{cite}
\usepackage{amsmath}
\usepackage{amsfonts}
\usepackage{bbm}
\usepackage[colorlinks=true, urlcolor=blue, linkcolor=blue, citecolor=red, pdfborder={0 0 0}]{hyperref}
\usepackage{subfig}
\usepackage{color}

\usepackage[absolute,overlay]{textpos} 

\def\msbar{\ensuremath{{\rm{\overline{MS}\;}}}} 
\newcommand{\ord}{{\ensuremath{\cal O}}}

\begin{document}

\begin{textblock*}{40mm}(1.0\textwidth,0.2\textheight)
\begin{flushright}
LPSC-15-180\\
IFJ PAN-IV-2015-9
\end{flushright}
\end{textblock*}

\title{Evolution kernels for parton shower Monte Carlo%
\thanks{Presented by A.~Kusina at Cracow Epiphany Conference, 8-10 January 2015}%
}
\author{A.~Kusina$^a$, O.\ Gituliar$^{^b}$, S.\ Jadach$^b$, M.\ Skrzypek$^b$
\address{$^a$ Laboratoire de Physique Subatomique et de Cosmologie,\\
              Universit\'e Grenoble-Alpes, CNRS/IN2P3,\\
              53 avenue des Martyrs, 38026 Grenoble, France}
\address{$^b$ Institute of Nuclear Physics, Polish Academy of Sciences,\\
              ul.\ Radzikowskiego 152, 31-342 Cracow, Poland}
}
\maketitle
\begin{abstract}
We report on re-calculation of the next-to-leading order DGLAP evolution kernels
performed in a scheme suited for Monte Carlo simulations of parton cascades (parton showers).
\end{abstract}

\section{Introduction}
In this contribution we report on the calculation of the next-to-leading
order (NLO) DGLAP evolution kernels that was started in
refs.~\cite{Jadach:2011kc,Gituliar:2014eba}. The inclusive DGLAP splitting
functions at NLO are known for a long time, they were first calculated in early
1980's~\cite{Floratos:1977au,Floratos:1978ny,Curci:1980uw,Furmanski:1980cm}.
Today also the next-to-next-to-leading order (NNLO) corrections are
known for over ten years~\cite{Moch:2004pa,Vogt:2004mw} so why the interest in NLO splitting
functions? The reason for this is the fact that all the above mentioned
calculations were performed in the \msbar scheme and by their nature are
inclusive. A direct motivation for the current work comes from Monte Carlo
simulations of parton cascades in form of parton shower (PS) generators.
In particular from the effort of the Cracow group that aims to include full NLO
corrections in the parton shower
simulations~\cite{Jadach:2010ew,Skrzypek:2011zw,Jadach:2011cr,Jadach:2015mza}.
This means not only corrections to the hard matrix element, as is done in
approaches like MC@NLO~\cite{Frixione:2002ik} and POWHEG~\cite{Nason:2004rx},
but also the full NLO corrections to the parton cascades.

The big difficulty with including NLO corrections in parton shower algorithms
is connected with the nature of parton shower generators. Specifically in PS
approach all the subsequent parton emissions need to be treated in fully exclusive
way, meaning that their 4-momenta are not integrated (including transverse degrees
of freedom). For comparison in a standard inclusive QCD calculations there are two
elements, Wilson coefficient (parton level hard matrix element) and parton distribution
functions (PDFs), that sum up all the above mentioned emissions in an inclusive
way.
Moreover if we go beyond LO in PS approach we need additional prescription to
avoid double counting of contributions in hard matrix element and in the shower.
This is referred to as the NLO matching and it was a major improvement for PS in
the last decade. It was originally done with MC@NLO method and shortly after the
POWHEG method was introduced. Recently an alternative solution from the Cracow
group is also available referred to as {\tt KrkNLO} method~\cite{Jadach:2015mza}.
This new solution is distinct from the earlier methods in a number of ways, one of
them being the use of different factorization scheme. Instead of using traditional \msbar
scheme that was designed for simplicity of inclusive calculations this method uses
a dedicated factorization scheme, referred to as Monte Carlo (MC) scheme, which was
designed especially for Monte Carlo simulations of parton cascades. The purpose of
introducing this scheme was not only NLO matching but rather
simplification of a more difficult task that is including NLO corrections in the
shower and maybe even later matching of the NLO shower with NNLO matrix element.
One of the necessary ingredients needed for this purpose are NLO evolution kernels
calculated in the MC scheme. We first calculated the real emission corrections to
the non-singlet NLO splitting functions~\cite{Jadach:2011kc} as these were required
for proof of concept of the methodology to include NLO corrections in the
shower~\cite{Jadach:2010ew,Skrzypek:2011zw}.
Recently we complemented the scheme by providing prescriptions for computing virtual
corrections and also calculated them for the non-singlet case~\cite{Gituliar:2014eba}.
Currently we are finishing the calculation of the NLO splitting functions
and in this contribution we are reporting on the progress.

\section{Monte Carlo friendly regularization scheme}
The regularization prescription for computing NLO splitting functions is one of the
ingredients of the Monte Carlo factorization scheme~\cite{Jadach:2015mza}.
It is based on the prescription of Curci, Furmnski and Petronzio (CFP)
introduced in~\cite{Curci:1980uw} during first NLO kernel calculation in $x$ space.
The most important features of the CFP scheme are
(i) the use of axial gauge;
and (ii) principal
value (PV) regularization for the axial gluon propagators [leading to infra red (IR)
singularities], namely:
\begin{equation}
g^{\mu\nu} - \frac{l^{\mu}n^{\nu}+l^{\nu}n^{\mu}}{ln}
\rightarrow
g^{\mu\nu} - \frac{l^{\mu}n^{\nu}+l^{\nu}n^{\mu}}{[ln]_{PV}},
\qquad\quad
\frac{1}{[ln]_{PV}} = \frac{ln}{(ln)^2+\delta^2(pl)},
\end{equation}
where $n$ is the axial gauge vector, $p$ is an external reference momentum
and $\delta$ is the PV geometrical regulator.
Unfortunately the use of PV prescription in this way leads to complicated patterns
for cancellation of singularities between real and virtual diagrams,
which is unacceptable if we want to use them for the construction of PS Monte Carlo.

\begin{figure}
\begin{center}
\captionsetup[subfloat]{labelformat=empty}
\subfloat[(NLO-qq-d-real)]{
\includegraphics[width=0.17\textwidth]{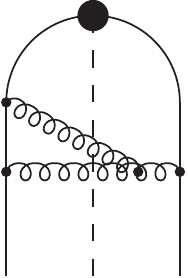}
}
\qquad
\subfloat[(NLO-qq-d-virt)]{
\includegraphics[width=0.17\textwidth]{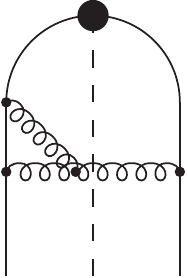}
}
\caption{Selected non-singlet diagrams contributing to the $P_{qq}$ kernel.}
\label{fig:NLO-qq-top-d}
\end{center}
\end{figure}

We illustrate it on the example of the non-singlet diagram NLO-qq-d displayed
in Fig.~\ref{fig:NLO-qq-top-d}. If calculated using CFP (\msbar) prescription
the virtual graph gives (we provide here only the singular terms):
\begin{equation}
\begin{split}
\Gamma^{\text{virt PV}}_{\text{NLO-qq-d}} &\sim
\frac{1}{\epsilon^3}\bigg[-p_{qq}(x)\bigg]
\\&+
\frac{1}{\epsilon^2}\bigg[
 p_{qq}(x)\bigg(-6I_0 - 2\ln(1-x) - \ln(x) + \frac{3}{2}\bigg)
 + (1-x)
                    \bigg]
\\&+
\frac{1}{\epsilon}\bigg[
 p_{qq}(x)\bigg(-2I_1 + 6I_0\ln(1-x) + 2I_0\ln(x)\bigg)
 + 6(1-x)I_0
                  \bigg],
\end{split}
\end{equation}
and the corresponding real graph reads:
\begin{equation}
\begin{split}
\Gamma^{\text{real PV}}_{\text{NLO-qq-d}} &\sim
\frac{1}{\epsilon^3}\bigg[p_{qq}(x)\bigg]
\\&+
\frac{1}{\epsilon^2}\bigg[
 p_{qq}(x)\bigg(2I_0\bigg)
 - (1-x)
                    \bigg]
\\&+
\frac{1}{\epsilon}\bigg[
 p_{qq}(x)\bigg(-2I_1 + 4I_0 - 2I_0\ln(1-x) + 2I_0\ln(x)\bigg)
 - 2(1-x)I_0
                  \bigg],
\end{split}
\end{equation}
where $p_{qq}(x)=\frac{1+x^2}{1-x}$ is the LO $qq$ kernel,
and $I_0$ and $I_1$ represent single and double logarithmic singularities
regulated by the PV regulator $\delta$:
\begin{equation}
\begin{split}
I_0 = & \int_0^1 dx\frac{1}{[x]_{PV}} = -\ln\delta +{\cal O}(\delta),
\\
I_1 = & \int_0^1 dx\frac{\ln x}{[x]_{PV}} = -\frac{1}{2} \ln^2\delta
-\frac{\pi^2}{24} +{\cal O}(\delta).
\end{split}
\end{equation}

We can see that higher order $\epsilon$ poles cancel between the corresponding real
and virtual graphs, which is probably impossible to implement in a MC shower program.

Because of this fact we have modified the CFP prescription by changing the way PV
regularization is applied. Instead of applying it only to the gluon propagators we
apply it to all the singularities in the plus variable, $l^+=nl/np$, see~\cite{Gituliar:2014eba}.
This results in simplification of the real-virtual cancellations which can be seen by
examining results for NLO-qq-d graphs in the new prescription (referred as NPV).
The virtual graph is given by:
\begin{equation}
\begin{split}
\Gamma^{\text{virt NPV}}_{\text{NLO-qq-d}} &\sim
\frac{1}{\epsilon^2}\bigg[
 p_{qq}(x)\bigg(-4I_0 - 2\ln(1-x) - \ln(x) + \frac{3}{2}\bigg)
                    \bigg]
\\&+
\frac{1}{\epsilon}\bigg[
 p_{qq}(x)\bigg(-6I_1 + 6I_0\ln(1-x) + 2I_0\ln(x)\bigg)
 + 4(1-x)I_0
                  \bigg],
\end{split}
\end{equation}
and the real graph is:
\begin{equation}
\begin{split}
\Gamma^{\text{real NPV}}_{\text{NLO-qq-d}} &\sim
\frac{1}{\epsilon}\bigg[
 p_{qq}(x)\bigg(2I_1 + 4I_0 - 2I_0\ln(1-x) + 2I_0\ln(x)\bigg)
                  \bigg].
\end{split}
\end{equation}
One can easily check that upon adding real and virtual graphs in both prescriptions
leads to the same final results.
However, from the MC point of view the situation in the NPV prescription is hugely improved.
There are no triple $\epsilon$ poles, the double poles occur only in the virtual graph and
the real one is free of them.%
    \footnote{More specifically, the $1/\epsilon^3$ terms have been replaced by $1/\epsilon\ln^2\delta$-type terms, regulated simultaneously by $\epsilon$ and $\delta$.
              }

Actually a general statement is valid.%
    \footnote{With some exceptions related to running of the coupling constant and CFP
    projection operators.}
In case of the NPV regularization: 
(i) $\epsilon^3$ poles in virtual and real graphs are absent (replaced by $1/\epsilon\ln^2\delta$-type structures), and (nearly) all $\epsilon^2$
poles originate from the virtual graphs alone,
(ii) all real emission graphs (contributing to NLO splitting functions) feature only
single $\epsilon$ poles,
(iii) sum of real and virtual graphs for the same topology reproduces the corresponding sum
calculated using CFP prescription.

There are some additional subtleties connected to the graphs contributing to the
running of the strong coupling (e.g. NLO-gg-f), these are partly discussed in ref.~\cite{Gituliar:2014eba}.

Using NPV regularization prescription is the first step on the way to the results
in the Monte Carlo scheme. For the MC shower simulations we need purely 4-dimensional
calculations and as we can see above the presented results still feature single 
(and partly double) $\epsilon$ poles.
This can be however avoided by introducing additional cut-off regularization for the
overall scale integration. It was already done in~\cite{Jadach:2011kc} when we calculated
real emission contributions to the non-singlet splitting functions
and won't be discussed here.

\section{Status of the calculation}
\begin{figure}
\captionsetup[subfloat]{labelformat=empty}
\subfloat[(NLO-gg-h) $-$ (NLO-gg-i)]{
\includegraphics[width=0.15\textwidth]{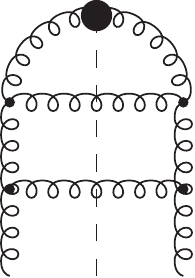}
\raisebox{30pt}{{\huge $-$}}
\includegraphics[width=0.13\textwidth]{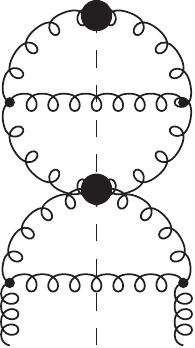}
\label{subfig:NLO-qg-top-hi}
}
\quad
\subfloat[(NLO-gg-b)]{
\includegraphics[width=0.15\textwidth]{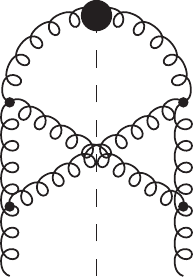}
\label{subfig:NLO-gg-top-b}
}
\quad
\subfloat[(NLO-gg-k)]{
\includegraphics[width=0.15\textwidth]{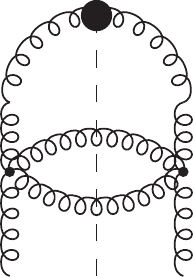}
\label{subfig:NLO-gg-top-k}
}
\quad
\subfloat[(NLO-gg-j)]{
\includegraphics[width=0.15\textwidth]{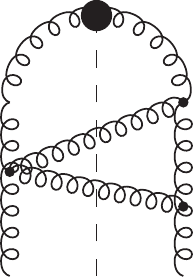}
\label{subfig:NLO-gg-top-j}
}
\newline
\subfloat[(NLO-gg-d)]{
\includegraphics[width=0.15\textwidth]{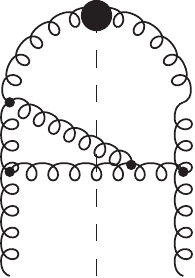}
\label{subfig:NLO-gg-top-d}
}
\quad
\subfloat[(NLO-gg-e)]{
\includegraphics[width=0.16\textwidth]{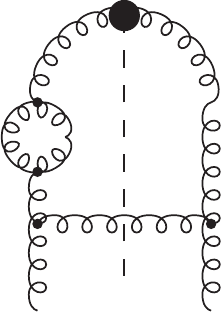}
\label{subfig:NLO-gg-top-e}
}
\quad
\subfloat[(NLO-gg-f)]{
\includegraphics[width=0.15\textwidth]{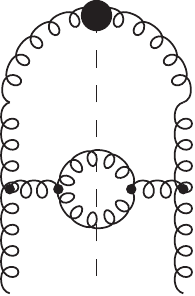}
\label{subfig:NLO-gg-top-f}
}
\quad
\subfloat[(NLO-gg-s1)]{
\includegraphics[width=0.15\textwidth]{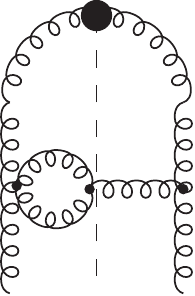}
\label{subfig:NLO-gg-top-s1}
}
\quad
\subfloat[(NLO-gg-s2)]{
\includegraphics[width=0.16\textwidth]{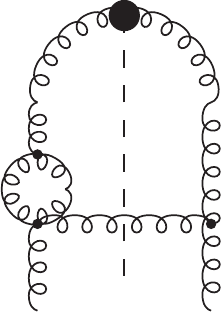}
\label{subfig:NLO-gg-top-s2}
}
\newline
\subfloat[(NLO-gg-hF) $-$ (NLO-gg-iF)]{
\includegraphics[width=0.15\textwidth]{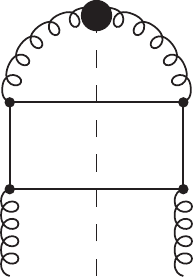}
\raisebox{30pt}{{\huge $-$}}
\includegraphics[width=0.13\textwidth]{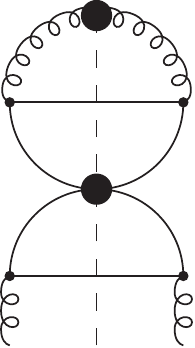}
\label{subfig:top-hi-Ferm}
}
\quad
\subfloat[(NLO-gg-bF)]{
\includegraphics[width=0.15\textwidth]{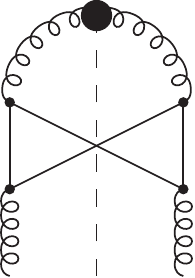}
\label{subfig:top-b-Ferm}
}
\newline
\subfloat[(NLO-gg-dF)]{
\includegraphics[width=0.15\textwidth]{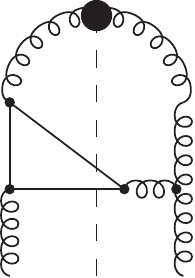}
\label{subfig:top-d-Ferm}
}
\quad
\subfloat[(NLO-gg-eF)]{
\includegraphics[width=0.16\textwidth]{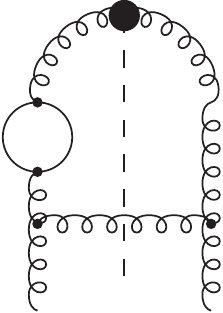}
\label{subfig:top-e-Ferm}
}
\quad
\subfloat[(NLO-gg-fF)]{
\includegraphics[width=0.15\textwidth]{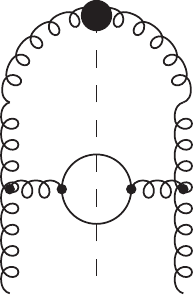}
\label{subfig:top-f-Ferm}
}
\caption{List of diagrams contributing to the $P_{gg}$ kernel. The graphs d,f,s1,dF and fF contribute in two versions, depending on the location of the cut.}
\label{fig:Pgg-contr}
\end{figure}

Currently we are finishing the calculation of the singlet splitting functions
using the prescription described in the previous section and introduced
in~\cite{Jadach:2011kc,Gituliar:2014eba}.
At the moment we have recalculated the full $P_{gg}$ and $P_{qg}$ kernel on the inclusive
level and we have obtained a perfect agreement with the literature.
Below we present the inclusive ``parton density'' (defined as in~\cite{Gituliar:2014mua})
for the case of $P_{gg}$. It is obtained as a sum of contributions from individual diagrams
in Fig.~\ref{fig:Pgg-contr}:
\begin{equation}
\begin{split}
\Gamma_{gg} &= 
\frac{C_A^2}{2\epsilon}
\left(\frac{\alpha_S}{2\pi}\right)^2
\Bigg\{
      \frac{1}{\epsilon}\bigg[
      p_{gg}(x)\left(\frac{11}{3}-8I_0-8\ln(1-x)-4\ln(x)\right)
\\&\qquad
      + \frac{44}{3}\left(\frac{1}{x}-x^2\right)
      - 12(1-x) + 8(1+x)\ln(x)
      \bigg]
      + 2p_{gg}(-x)S_2(x)
\\&\qquad
   +p_{gg}(x)\left(
   \ln^2(x) - 4\ln(x)\ln(1-x) - \frac{\pi^2}{3} + \frac{67}{9}
   \right)
      + 4(1+x)\ln^2(x)
\\&\qquad
   - \left(\frac{44}{3}x^2-\frac{11}{3}x+\frac{25}{3}\right)\ln(x)
   + \frac{27}{2}(1-x)
   + \frac{67}{9}\Big(x^2-\frac{1}{x}\Big)
\Bigg\}
\\& +
\frac{C_AT_f}{2\epsilon}
\left(\frac{\alpha_S}{2\pi}\right)^2
\bigg[ \left(-\frac{8}{3}\frac{1}{2\epsilon} - \frac{20}{9}\right)p_{gg}(x)
+\frac{26x^2}{9} - 2x + 2 - \frac{26}{9x}
\\&\qquad
- \frac{4}{3}(1+x)\ln(x)
\bigg]
\\&+
\frac{C_FT_f}{2\epsilon}
\left(\frac{\alpha_S}{2\pi}\right)^2
\bigg[ \frac{1}{2\epsilon}\left(4x - 4 - \frac{16}{3}\frac{1-x^3}{x} - 8(1+x)\ln(x)\right)
\\&\qquad
+ \frac{20x^2}{3} + 8x - 16 + \frac{4}{3x} - 2(1+x)\ln^2(x) - (10x+6)\ln(x)
\bigg],
\end{split}
\end{equation}
where:
\begin{equation}
\begin{split}
p_{gg}(x) &= \frac{(1-x+x^2)^2}{x(1-x)},
\\
S_2(x) &= \int_{\frac{x}{1+x}}^{\frac{1}{1+x}} \frac{dz}{z} \ln\Big(\frac{1-z}{z}\Big).
\end{split}
\end{equation}
The kernel is now easily extracted from $\Gamma_{gg}$ by taking twice its
residue (twice as we use $n=4+2\epsilon$); additionally we exclude factor
$\left(\frac{\alpha_S}{2\pi}\right)^2$ which is the usual convention, then:
\begin{equation}
\begin{split}
P_{gg}^{(1)} &= 2\,\text{Res}\left(\Gamma_{gg}\right) / \left(\frac{\alpha_S}{2\pi}\right)^2
\\&=
C_A^2
\bigg[
   p_{gg}(x)\left(
   \ln^2(x) - 4\ln(x)\ln(1-x) - \frac{\pi^2}{3} + \frac{67}{9}
   \right)
      + 2p_{gg}(-x)S_2(x)
\\&\qquad\qquad
   + 4(1+x)\ln^2(x)
   - \left(\frac{44}{3}x^2-\frac{11}{3}x+\frac{25}{3}\right)\ln(x)
\\&\qquad\qquad
   + \frac{27}{2}(1-x) + \frac{67}{9}\Big(x^2-\frac{1}{x}\Big)
\bigg]
\\& +
C_AT_f
\bigg[ - \frac{20}{9}p_{gg}(x)
+\frac{26x^2}{9} - 2x + 2 - \frac{26}{9x} - \frac{4}{3}(1+x)\ln(x)
\bigg]
\\&+
C_FT_f
\bigg[ \frac{20x^2}{3} + 8x - 16 + \frac{4}{3x} - 2(1+x)\ln^2(x) - (10x+6)\ln(x)
\bigg].
\end{split}
\end{equation}
This result reproduces the well known $P_{gg}$ kernel, that can be found
in ref.~\cite{Ellis:1996nn} or in the original Furmanski
and Petronzio paper~\cite{Furmanski:1980cm}.
The literature provides the single $\epsilon$ terms defining the splitting
functions but some of the double pole terms from the inclusive densities can
also be cross-checked with the results in~\cite{Hei98}.

\begin{figure}
\captionsetup[subfloat]{labelformat=empty}
\subfloat[(NLO-qg-h1) $-$ (NLO-qg-i1)]{
\includegraphics[width=0.15\textwidth]{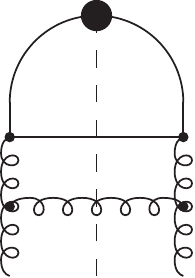}
\raisebox{30pt}{{\huge $-$}}
\includegraphics[width=0.13\textwidth]{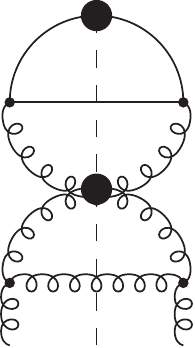}
\label{subfig:NLO-qg-top-h1i1}
}
\qquad
\subfloat[(NLO-qg-h2) $-$ (NLO-qg-i2)]{
\includegraphics[width=0.15\textwidth]{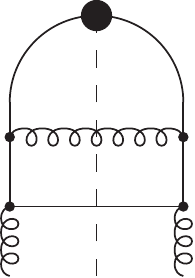}
\raisebox{30pt}{{\huge $-$}}
\includegraphics[width=0.13\textwidth]{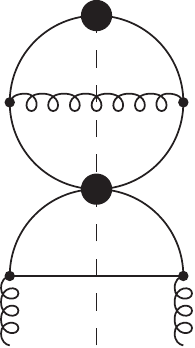}
\label{subfig:NLO-qg-top-h2i2}
}
\newline
\subfloat[(NLO-qg-b)]{
\includegraphics[width=0.15\textwidth]{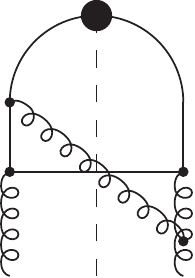}
\label{subfig:NLO-qg-top-b}
}
\quad
\subfloat[(NLO-qg-d1)]{
\includegraphics[width=0.15\textwidth]{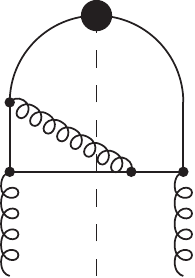}
\label{subfig:NLO-qg-top-d1}
}
\quad
\subfloat[(NLO-qg-d2)]{
\includegraphics[width=0.15\textwidth]{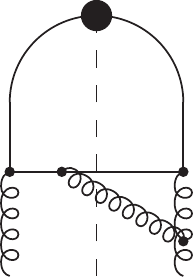}
\label{subfig:NLO-qg-top-d2}
}
\quad
\subfloat[(NLO-qg-f)]{
\includegraphics[width=0.15\textwidth]{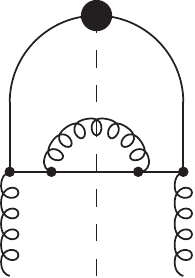}
\label{subfig:NLO-qg-top-f}
}
\quad
\subfloat[(NLO-qg-e)]{
\includegraphics[width=0.15\textwidth]{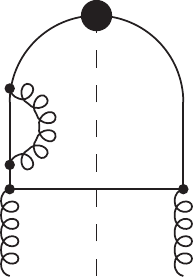}
\label{subfig:NLO-qg-top-e}
}
\caption{List of diagrams contributing to the $P_{qg}$ kernel. The graphs d1, d2 and f contribute in two versions, depending on the location of the cut.}
\label{fig:Pqg-contr}
\end{figure}

Similarly for the case of $P_{qg}$ the inclusive density is given by
(the contributing diagrams are listed in Fig.~\ref{fig:Pqg-contr}):
\begin{equation}
\begin{split}
\Gamma_{qg} &= 
\frac{C_FT_f}{2\epsilon}
\left(\frac{\alpha_S}{2\pi}\right)^2
\bigg\{
\frac{1}{2\epsilon}\bigg[p_{qg}(x)\Big(-16I_0 - 8\ln(1-x) - 8\ln(x) + 12\Big)
\\&\qquad
- 4(1-2x)\ln(x) - 8x + 2\bigg]
+ p_{qg}(x)\bigg[
               2\ln^2(1-x) + 2\ln^2(x)
\\&\qquad
               - 4\ln(x)\ln(1-x) - 4\ln(1-x) + 4\ln(x) - \frac{2\pi^2}{3} + 10
           \bigg]
\\&\qquad
  - (1-2x)\ln^2(x) - (1-4x)\ln(x) + 4\ln(1-x) - 9x + 4
\bigg\}
\\&+
\frac{C_AT_f}{2\epsilon}
\left(\frac{\alpha_S}{2\pi}\right)^2
\bigg\{
\frac{1}{2\epsilon}\bigg[ p_{qg}(x)\Big(-8\ln(1-x) + \frac{62}{3}\Big)
      + \frac{28x}{3} - \frac{74}{3} - \frac{16}{3x}
\\&\qquad
      - (32x+8)\ln(x)
                   \bigg]
+ p_{qg}(x)\bigg[
               -2\ln^2(1-x) - \ln^2(x) + 4\ln(1-x)
\\&\qquad
               + \frac{44}{3}\ln(x) + \frac{\pi^2}{3} - \frac{218}{9}
           \bigg]
   + 2p_{qg}(-x)S_2(x) + \frac{14x}{9} + \frac{40}{9x} + \frac{182}{9}
\\&\qquad
   - (8x+2)\ln^2(x) - \Big(\frac{38}{3}-\frac{136x}{3}\Big)\ln(x)
   - 4\ln(1-x)
\bigg\},
\end{split}
\end{equation}
where $p_{qg}(x) = x^2 + (1-x)^2$ is the corresponding LO kernel.
The resulting splitting function is equal to:
\begin{equation}
\begin{split}
P_{qg}^{(1)} &= 2\,\text{Res}\left(\Gamma_{qg}\right) / \left(\frac{\alpha_S}{2\pi}\right)^2
\\& = 
C_FT_f
\bigg[
p_{qg}(x)\bigg(
               2\ln^2(1-x) + 2\ln^2(x) - 4\ln(x)\ln(1-x) - 4\ln(1-x)
\\&\qquad\qquad\qquad\qquad
               + 4\ln(x) - \frac{2\pi^2}{3} + 10
         \bigg)
\\&\qquad\qquad
  - (1-2x)\ln^2(x) - (1-4x)\ln(x) + 4\ln(1-x) - 9x + 4
\bigg]
\\&+
C_AT_f
\bigg[
p_{qg}(x)\bigg(
               - 2\ln^2(1-x) - \ln^2(x) + 4\ln(1-x) + \frac{44}{3}\ln(x)
\\&\qquad\qquad
               + \frac{\pi^2}{3} - \frac{218}{9}
          \bigg)
   + 2p_{qg}(-x)S_2(x) + \frac{14x}{9} + \frac{40}{9x} + \frac{182}{9}
\\&\qquad\qquad
   - (8x+2)\ln^2(x) - \Big(\frac{38}{3}-\frac{136x}{3}\Big)\ln(x)
   - 4\ln(1-x)
\bigg],
\end{split}
\end{equation}
which agrees with the literature~\cite{Furmanski:1980cm,Ellis:1996nn}.

The double $\epsilon$ pole terms in the inclusive densities $\Gamma$ originate from two
sources. First is building up of the running coupling constant, the second is artificial
and connected with the subtraction defined through the projection operator
($\mathbbm{P}$ in CFP~\cite{Curci:1980uw}), entering via countergraphs like NLO-qg-i1 or NLO-gg-i.
In the inclusive calculations the higher
$\epsilon$ poles are not important as splitting functions do not depend on them (they
are defined by the residue -- single pole part), making this kind of projection operator
perfectly well suited for the inclusive computation. However, if we want to use the
unintegrated distributions (exclusive version of $\Gamma$) in the Monte Carlo program
this will become a problem. We can solve it by using a dedicated Monte Carlo factorization
scheme~\cite{Jadach:2015mza,Jadach:2010ew} which redefines projection operators, however,
this is beyond the scope of this work and we will not investigate it here.
Let us just mention that similar idea has been investigated by Oliveira
{\em et al.}~\cite{deOliveira:2013maa,deOliveira:2013iya} in the context of physical
factorization scheme allowing for better convergence of perturbative series.

\begin{figure}
\captionsetup[subfloat]{labelformat=empty}
\subfloat[(NLO-gq-h1) $-$ (NLO-gq-i1)]{
\includegraphics[width=0.15\textwidth]{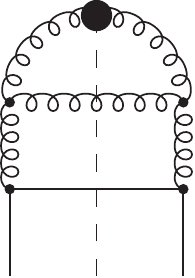}
\raisebox{30pt}{{\huge $-$}}
\includegraphics[width=0.13\textwidth]{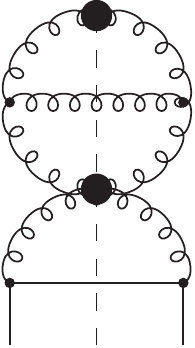}
\label{subfig:NLO-gq-top-h1i1}
}
\qquad
\subfloat[(NLO-gq-h2) $-$ (NLO-gq-i2)]{
\includegraphics[width=0.15\textwidth]{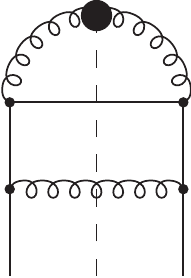}
\raisebox{30pt}{{\huge $-$}}
\includegraphics[width=0.13\textwidth]{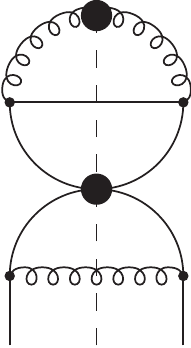}
\label{subfig:NLO-gq-top-h2i2}
}
\quad
\subfloat[(NLO-gq-b)]{
\includegraphics[width=0.15\textwidth]{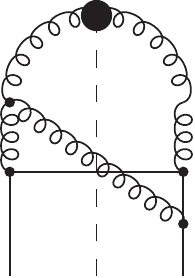}
\label{subfig:NLO-gq-top-b}
}
\newline
\subfloat[(NLO-gq-d1)]{
\includegraphics[width=0.15\textwidth]{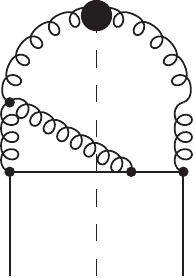}
\label{subfig:NLO-gq-top-d1}
}
\quad
\subfloat[(NLO-gq-d2)]{
\includegraphics[width=0.15\textwidth]{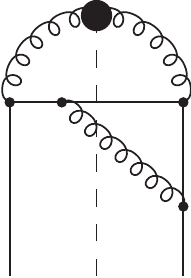}
\label{subfig:NLO-gq-top-d2}
}
\quad
\subfloat[(NLO-gq-f)]{
\includegraphics[width=0.15\textwidth]{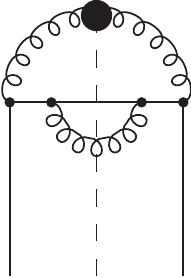}
\label{subfig:NLO-gq-top-f}
}
\quad
\subfloat[(NLO-gq-e1)]{
\includegraphics[width=0.15\textwidth]{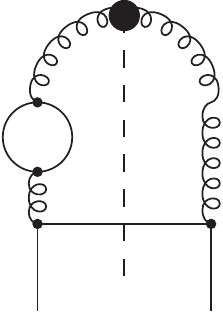}
\label{subfig:NLO-gq-top-e1}
}
\quad
\subfloat[(NLO-gq-e2)]{
\includegraphics[width=0.15\textwidth]{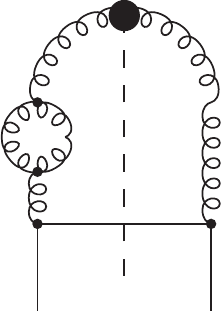}
\label{subfig:NLO-gq-top-e2}
}
\caption{List of diagrams contributing to the $P_{gq}$ kernel. The graphs d1, d2 and f contribute in two versions, depending on the location of the cut.}
\label{fig:Pgq-contr}
\end{figure}

\begin{figure}
\captionsetup[subfloat]{labelformat=empty}
\begin{center}
\subfloat[(NLO-qqb-h1) $-$ (NLO-qqb-i1)]{
\includegraphics[width=0.15\textwidth]{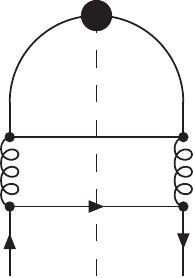}
\raisebox{30pt}{{\huge $-$}}
\includegraphics[width=0.13\textwidth]{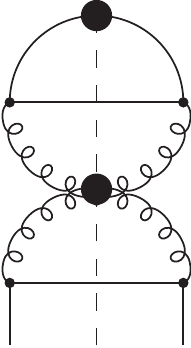}
\label{subfig:NLO-qqbar-top-hi}
}
\qquad
\subfloat[(NLO-qqb-b)]{
\includegraphics[width=0.15\textwidth]{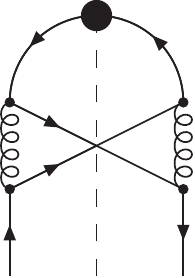}
\label{subfig:NLO-qqbar-top-b}
}
\caption{List of diagrams contributing to the $P_{q\bar{q}}$ kernel.}
\label{fig:Pqqbar-contr}
\end{center}
\end{figure}

Currently we are working to complete the calculation for $P_{gq}$ and $P_{q\bar{q}}$ kernel.
Diagrams contributing to these calculations can be seen in Figs.~\ref{fig:Pgq-contr}
and~~\ref{fig:Pqqbar-contr}.
Computation for $P_{gq}$ is analogical to the one for $P_{qg}$.
In case of $P_{q\bar{q}}$ the first (Born) level contributions appear only at order
$\ord(\alpha_S^2)$ making this calculation very easy.

\section{Summary}
We have reported on the progress of re-calculation of the NLO splitting function
using new NPV regularization prescription~\cite{Gituliar:2014eba}. We have already
calculated the $P_{gg}$ and $P_{qg}$ (and earlier $P_{qq}$) NLO splitting functions
and we have reproduced the know inclusive results from the literature.
The calculations for the $P_{gq}$ splitting function is already advanced and the partial
results we have obtained so far are also in agreement with the literature.

The inclusive results presented here are {\em not} our main interest and serve us mainly as a
cross-check of our calculations. The main interest for us are the unintegrated
distributions in the new scheme that will be better suited for the Monte Carlo
simulations. These distributions will be presented elsewhere.

%
%

\section*{Acknowledgments}
This work has been partly supported by the Polish National Science Center 
grants DEC-2011/03/B/ST2/02632 and UMO-2012/04/M/ST2/00240.
The work of O.G. has been supported by Narodowe Centrum Nauki with the 
Sonata Bis grant DEC-2013/10/E/ST2/00656. 

\bibliographystyle{utphys}
\bibliography{radcor}

\end{document}